\input{aipcheck}

\documentclass[
    ,final            
  ]
  {aipproc}

\layoutstyle{6x9}

\begin{document}

\title{Neutral Pion Double Longitudinal Spin Asymmetry in Proton-Proton Collisions at $\sqrt{s}=200$ GeV Using the PHENIX Detector}


\author{Kieran Boyle on behalf of the PHENIX collaboration}{
  address={Stony Brook University, Stony Brook, NY, 11794}
}

\begin{abstract}
 New results from polarized p-p collisions at $\sqrt{s}=200$ GeV
 of double longitudinal spin asymmetry in $\pi^0$ production using
 the PHENIX detector in the 2005 RHIC run are presented.  Both
 positive and negative maximal gluon polarization scenarios are
 inconsistent with the data.  The data is consistent with small $|\Delta
 g|$, including $\Delta g=0$.
\end{abstract}

\maketitle


\section{Introduction}

In the past 20 years, polarized DIS fixed target experiments have
established that the quarks carry 20-30\% of the nucleon spin. The
remainder must be due to gluon spin and the orbital angular momentum
of the quarks and gluons.  In proton-proton collisions, we can
examine the gluon spin contribution through lowest order gluon-gluon
and gluon-quark interactions. In this report, we focus on the double
longitudinal spin asymmetry, $A_{LL}$, of $\pi^{0}$ production for
$\eta<|0.35|$.

We define $A_{LL}$ as
\begin{equation}
A_{LL}=\frac{\sigma_{++} - \sigma_{+-}}{\sigma_{++} + \sigma_{+-}},
\label{A_LL crosssection}
\end{equation}
where $\sigma_{++}$ ($\sigma_{+-}$) is the measured cross section,
in this case of the $\pi^0$, with same (opposite) helicity beams.
Rewriting this in terms of our measured quantities, we get
\begin{equation}
A_{LL}=\frac{1}{|P_1||P_2|}\frac{N_{++} - RN_{+-}}{N_{++} +
RN_{+-}},\quad R=\frac{L_{++}}{L_{+-}}. \label{A_LL RL}
\end{equation}
where $L$ is the integrated luminosity, $R$ is the relative
luminosity, $P_1$ and $P_2$ are beam polarizations and N is the
particle yield.

\section{Measurement}

In 2005, the absolute polarization in RHIC was measured with a
polarized hydrogen gas jet target \cite{RHIC:poljet}. A
proton-Carbon Coulomb Nuclear Interference (CNI) polarimeter
\cite{RHIC:CNI} was used on a fill-by-fill basis for measuring the
relative polarization. For 2005, the average polarization was 47\%,
with 20\% relative error per beam. This gives a 40\% scaling
uncertainty in $A_{LL}^{\pi^0}$.

The stable beam polarization direction in RHIC is vertical.  For
longitudinal spin measurements, the beams pass through spin rotators
to align the polarization vector with beam momentum. Polarization
direction in PHENIX is measured by observing the single transverse
spin asymmetry of forward neutron production \cite{pi0:2003}.  In
2005, both beams were found to be above 98\% longitudinally
polarized at PHENIX.

To measure relative luminosity, we use beam-beam counters
\cite{pi0:2003}. In PHENIX for the 2005 run, uncertainty in relative
luminosity was $\Delta R = 1.0\times10^{-4}$, corresponding to a
$\Delta A_{LL}|_{_{R}}=2.3\times10^{-4}$.

The electromagnetic calorimeter of PHENIX central arms
\cite{PHENIX:detector} and a high $p_T$ photon trigger are used to
measure $\pi^0$ yield. The $\pi^0$ cross section for the 2005 run
was found to be in good agreement with results from previous runs
\cite{pi0:cross}. NLO QCD calculations are consistent with $\pi^0$
cross section, which gives us confidence when comparing our final
$A_{LL}$ results with theoretical expectations for different values
of $\Delta G$.

\section{Asymmetry Calculation}

Asymmetry is calculated using Eq. \ref{A_LL RL} for the diphoton
invariant-mass range of $\pm25$ MeV/c$^2$ around the $\pi^0$ peak
($A_{LL}^{\pi^0+BG}$) and in two 50 MeV/c$^2$ mass regions on either
side ($A_{LL}^{BG}$).  Fitting the mass peak, we obtain $w_{BG}$,
the fraction of background in the peak region.  To obtain the
$A_{LL}^{\pi^0}$, we use
\begin{equation}
A_{LL}^{\pi^0}=\frac{A_{LL}^{\pi^0+BG} -
w_{BG}A_{LL}^{BG}}{1-w_{BG}}. \label{A_LL crosssection}
\end{equation}

To evaluate systematic error between bunches and fills, we use bunch
shuffling \cite{pi0:2003}.  For the 2005 run, bunch-to-bunch and
fill-to-fill systematic uncertainty is negligible with our current
statistics.

\section{$\pi^{0} A_{LL}$ Results}

\begin{table}
\begin{tabular}{lccc}
\hline
  & \tablehead{1}{c}{b}{<Polarization> (\%)}
  & \tablehead{1}{c}{b}{Integrated Luminosity}
  & \tablehead{1}{c}{b}{Figure of Merit (P$^4$L)}  \\
\hline
2003 RHIC run & 27 & 220 nb$^{-1}$ & 1.17 nb$^{-1}$ \\
2004 RHIC run & 40 & 75 nb$^{-1}$ & 1.92 nb$^{-1}$ \\
2005 RHIC run & 47 & 1.8 pb$^{-1}$ & 87.8 nb$^{-1}$ \\
\hline
\end{tabular}
\caption{Comparison between RHIC runs.  The 2005 run was a
substantial increase in figure of merit over previous runs due to
increases in both polarization and integrated luminosity.}
\label{table_FoM}
\end{table}

\begin{figure}
  \includegraphics[height=.25\textheight]{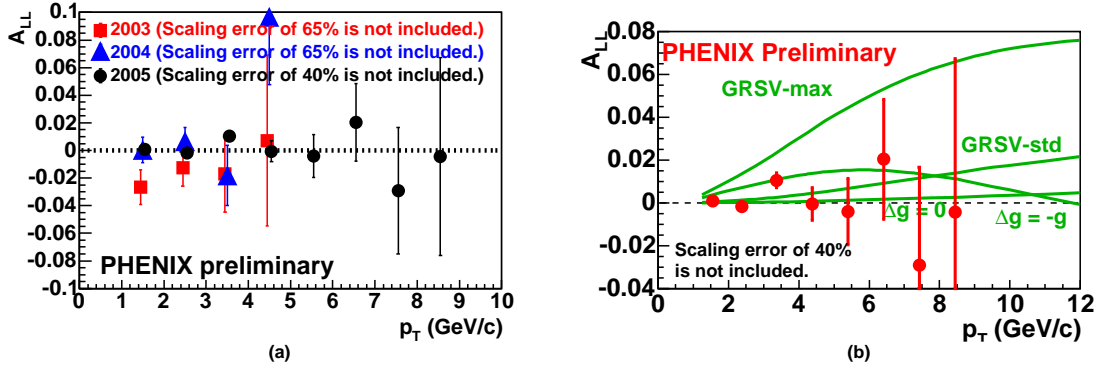}
  \caption{(a)  $A_{LL}^{\pi^0}$ results from RHIC runs in 2003, 2004 and
  2005 are consistent.  (b)  $A_{LL}^{\pi^0}$ from 2005 RHIC run
  with four theory curves.  Data excludes maximal cases of $\Delta
  $G $=$ G (GRSV-max) and $\Delta $G $=-$G.}
  \label{fig:ALL_plot}
\end{figure}

PHENIX has previously obtained $A_{LL}^{\pi^0}$ for the 2003 RHIC
run \cite{pi0:2003} and the 2004 RHIC run \cite{pi0:2004}.  These
results are shown in Fig. \ref{fig:ALL_plot} (a) along with the 2005
run result. Table \ref{table_FoM} lists the average polarization,
integrated luminosity (used in $A_{LL}^{\pi^0}$ analysis), and
figure of merit (P$^4$L) for the three RHIC runs. The 2005 RHIC run
showed a substantial gain in figure of merit, due to increase in
polarization and luminosity at RHIC. Results from the three runs are
consistent.

Figure \ref{fig:ALL_plot} (b) shows the 2005 run result with 4
theory curves \cite{pi0:GRSV}.  Confidence level between our data
and GRSV standard range from 17.1-21.7\%, accounting for 40\%
scaling uncertainty from beam polarization.  No uncertainty from
theory is considered. Three other theory curves, GRSV-max ($\Delta$G
$=$ G), $\Delta$G $=-$G, and $\Delta$G $=0$, all with input scale
$Q^2 = 0.4$ GeV$^2$, are also shown. Confidence level ranges for
comparing our data with each are 0.0-0.0\%, 0.0-0.7\% and
16.7-18.4\%, respectively.  From these, we conclude that our data
disagree with maximal gluon polarization ($\Delta$G $=$ G or
$\Delta$G $ = -$G) but are consistent with GRSV standard and
$\Delta$G $ = 0$.

\section{Summary}

In the 2005 RHIC run, $A_{LL}^{\pi^0}$ was measured at mid-rapidity
and $\sqrt{s}=200$ GeV with the PHENIX detector.  The figure of
merit was significantly higher than previous runs due to increases
in polarization and integrated luminosity.  Results are consistent
with GRSV standard and $\Delta $G $= 0$, but disagree with positive
and negative maximal $\Delta$G.





\bibliographystyle{aipproc}  

\bibliography{panic}


\end{document}